\documentstyle[prl,aps,epsf,floats]{revtex}
\begin{document}
\draft
\twocolumn[\hsize\textwidth\columnwidth\hsize\csname @twocolumnfalse\endcsname
\title {The pseudogap and photoemission spectra 
in the attractive Hubbard model}
\author{P.\,E.\,Kornilovitch$^1$ and Bumsoo Kyung$^2$}
\address{$^1$Blackett Laboratory, Imperial College,
Prince Consort Road, London SW7 2BZ, UK\\
$^2$Max-Planck-Institut f\"ur Physik komplexer Systeme, 
N\"othnitzer str. 38, D-01187, Dresden, Germany}
\date{\today}
\maketitle
\begin{abstract}

Angle-resolved photoemission spectra are calculated 
microscopically for the two-dimensional attractive Hubbard
model. A system of self-consistent $T$-matrix equations are  
solved numerically in the real-time domain. The single-particle 
spectral function has a two-peak structure resulting from
the presense of bound states. The spectral function is suppressed
at the chemical potential, leading to a pseudogap-like behavior.
At high temperatures and densities the pseudogap diminishes
and finally disappears; these findings are similar to experimental 
observations for the cuprates.

\end{abstract}
\vskip2pc]
\narrowtext

Real-space pairing \cite{Alexandrov} is the simplest physical 
idea that enables
one to explain the pseudogap phenomenon observed for the normal
state of high-$T_c$ superconductors (HTSC) \cite{Takigawa,Williams_one,Loram,Ito,Batlogg,Rotter,Homes,Loeser,Nemetschek,Tjernberg,Williams,Puchkov}.
At low temperatures and densities carriers are paired in 
weakly overlapping bound states separated
from the single-particle band by a binding energy of the order
of a few hundred degrees. The chemical potential is located between
the two bands thereby reducing the intensity of the low-energy
single-particle (photoemission, specific heat, tunneling), spin
(susceptibility, nuclear relaxation rate), 
and particle-hole (optical conductivity)
processes. At temperatures of the order of the binding energy,
carriers become unbound and restore the Fermi-liquid behavior.
The layered structure of the cuprates
supports the scenario, since reduced dimensionality 
favors pairing. Also, the phenomenology of a charged
Bose gas can be quite successfully used in the explanation of a number
of normal and superconducting properties of HTSC 
\cite{Alexandrov,Uemura,Alexandrov_two}.

It is therefore important to study model systems with pairing,
such as the two-dimensional attractive Hubbard model. The
simplicity of the model allows one to separate the net effect of
the attractive interaction from the complications related to the 
origin of the pairing mechanism and to the complicated dependence
of the effective potential on microscopic parameters.
One successful application of the attractive Hubbard model
to the physics of cuprates is due to Randeria and co-workers
\cite{Randeria_one,Randeria_two}. Using the Quantum Monte Carlo
method, they found a significant reduction of static spin
susceptibility and nuclear relaxation rate at low temperatures
and intermediate couplings. Recently, Vilk {\em et al.} 
\cite{Vilk} found a pseudogap in the spectral function 
of the attractive Hubbard model using Monte Carlo simulations
and the maximum-extropy technique. Thus, it was demonstrated 
that real-space pairing can account for some unusual 
properties of HTSC.

The self-consistent $T$-matrix approximation 
\cite{Kanamori,Fetter,Kadanoff} provides another method
for studying dynamic properties of the attractive Hubbard
model. This approach is based on the low-density approximation 
to fermion systems due to Galitskii \cite{Galitskii}, 
which becomes exact in the limit of vanishing density. 
However, the resulting system of self-consistent 
integral equations is not easy to solve. Analytical
treatment is very difficult, although it is sometimes attempted  
\cite{Kagan,Letz}. The full numerical solution of the
equations is required. In the previous
numerical studies of the problem \cite{Wolfle,Micnas} 
the equations were solved in imaginary times and
results were then continued numerically to real times.  
In these papers relatively small couplings were studied
and pseudogap features were found only at large momenta.  

In our previous paper \cite{Kyung} we formulated and solved 
numerically the $T$-matrix equations for the two-dimensional 
attractive Hubbard model for real times, thereby avoiding the
necessity for analytical continuation. There, we focused
on two-particle properties --- primarily on the binding
energy of pairs and its dependence on the particle
density. In this paper we would like to discuss single-particle
dynamics, in particular the single-particle spectral function
which is directly related to angle-resolved photoemission
spectra (ARPES). We find a clear pseudogap behavior of the ARPES 
at small momenta ${\bf k}$, low densities $n$, and low temperatures
$T$. With increasing $n$ and $T$, the pseudogap disappears, 
in accordance with experimental observations for the cuprates
\cite{Ding}.

The two-dimensional attractive
Hubbard model is defined by the Hamiltonian
\begin{equation}
H = \sum_{{\bf k}\sigma} (\varepsilon_{\bf k} -\mu)
c^{\dagger}_{{\bf k}\sigma} c_{{\bf k}\sigma} -
\frac{|U|}{N} \sum_{\bf k p q}
c^{\dagger}_{{\bf k} \uparrow} c_{{\bf k+q} \uparrow}
c^{\dagger}_{{\bf p} \downarrow} c_{{\bf p-q} \downarrow} ,
\label{one}
\end{equation}
written in standard notation. 
$\varepsilon_{\bf k}=-2t \, (\cos{k_x}+\cos{k_y})$ is the
bare single-particle spectrum, $|U|$ is the coupling strength, and
$N$ is the total number of sites in the system. The chemical 
potential $\mu$ determines the average particle density $n$. 
We regard Eq. (\ref{one}) as a phenomenological model for
the low-density system of {\em holes} in the normal state
of HTSC.

In the low-density
limit $n \ll 1$, one can make use of the small gas parameter and
select only ladder diagrams in a diagrammatic representation
of the $T$-matrix \cite{Fetter,Kadanoff},
which leads to the expression
\begin{equation}
T({\bf q}, \omega) = \frac{-|U|} {1-|U| \displaystyle{
\int \frac{d\omega_1}{2\pi} 
\frac{B({\bf q}, \omega_1)}{\omega-\omega_1} + 
i \frac{|U|}{2} B({\bf q}, \omega) }} ,       
\label{two}
\end{equation}
where
\begin{equation}
B({\bf q}, \omega) \!=\! \frac{-1}{N} \! \sum_{\bf k'}
\!\! \int \!\! \frac{d\omega_1}{2\pi} A({\bf k'}\!, \omega_1)
A({\bf q\!-\!k'}\!, \omega\!-\!\omega_1)  \tanh{\!\frac{\beta
\omega_1}{2}} ,
\label{three}
\end{equation}
where $A({\bf k}, \omega)$ is the single-particle spectral
function and $\beta=(k_B T)^{-1}$ is the
inverse absolute temperature.
The real and imaginary parts of the self-energy
$\Sigma'$ and $\Sigma''$ are expressed via $T = T' + i T''$
as follows:
\begin{displaymath}
\Sigma'({\bf k}, \omega) = \frac{1}{N} \sum_{\bf q}
\int \frac{d \omega_1}{2\pi} A({\bf q-k}, \omega_1) \times
\end{displaymath}
\vspace{-0.5cm}
\begin{equation}
\times \left[ f_F(\omega_1) T'({\bf q}, \omega+\omega_1) +
\int \frac{d \omega_2}{\pi}
\frac{f_B(\omega_2) T''({\bf q}, \omega_2)}{\omega_2-\omega_1-\omega}
\right] ,
\label{four}
\end{equation}
\begin{displaymath}
\Sigma''({\bf k}, \omega) = \frac{1}{N} \sum_{\bf q}
\int \frac{d \omega_1}{2\pi} A({\bf q-k}, \omega_1) \times
\makebox[2.cm]{}
\end{displaymath}
\vspace{-0.3cm}
\begin{equation}
\times T''({\bf q}, \omega+\omega_1)
\left[ f_F(\omega_1) + f_B(\omega+\omega_1) \right] ,
\label{five}
\end{equation}
where $f_{F,B}(\omega)=[\exp(\beta \omega) \pm 1]^{-1}$ are Fermi- and 
Bose- functions respectively. Finally, the self-energy 
determines the spectral function as
\begin{equation}
A({\bf k}, \omega) = \frac{-\, 2\, \Sigma''({\bf k}, \omega)}
{[\omega - (\varepsilon_{\bf k}-\mu)-\Sigma'({\bf k}, \omega)]^2
+ [\Sigma''({\bf k}, \omega)]^2} .
\label{six}
\end{equation}
The integrals with singular kernels in Eqs.(\ref{two}) and
(\ref{four}) are understood in the principal-value sense.
The set of equations (\ref{two})-(\ref{six}) is to be solved
self-consistently for given $|U|$, $\mu$, and temperature
$T$; then the particle density is given by
$n = 2 \, N^{-1} \sum_{\bf k} \int \frac{d \omega}{2\pi}
A({\bf k},\omega) f_F(\omega)$.
Usually, a self-consistent solution is obtained iteratively,
starting from a guessed form of $A({\bf k},\omega)$ and using
the fast-Fourier-transform algorithm to calculate 
momentum-frequency sums \cite{Wolfle,Micnas}. In our
calculations we used a $64 \times 64$ lattice and a uniform
mesh of 512 points in the frequency interval 
$-20\,t < \omega < 30\, t$. The convergence of the iterative
process is the major problem of the method, which puts
limitations on the values of model parameters for which
a self-consistent solution can be obtained. The convergence
deteriorates for large $|U|$ and $n$ and low $T$. The physically 
interesting values of $|U|$ start at $\sim 6\,t$, when the binding
energy of the pairs is of the order of $t$. In this work, 
$|U|=8\,t$ is used. For this coupling, iterations converge down to
$T=0.3 \, t$ for very low densities $n < 0.03$, and up to
$n\sim 0.20$ for a high temperature $T=1.0\,t$. 

Once a self-consistent solution is obtained, the intensity
of the photoemission process is simply
\begin{equation}
I({\bf k}, \omega) = I_0({\bf k}) \, A({\bf k}, \omega) f_F(\omega) ,
\label{eight}
\end{equation}
where $I_0({\bf k})$ involves the electron-photon matrix element, and
is frequency-independent. Eq. (\ref{eight}) is approximate, for
a discussion of its validity see, e.g., \cite{Campuzano}.
In the following we set $I_0({\bf k}) = 1$.

In analysing the numerical results to be presented below, 
it is useful to keep in mind
the exactly solvable atomic limit ($t=0$) of the Hubbard model:
\begin{equation}
\frac{1}{2 \pi} A({\bf k}, \omega) = \frac{n}{2}
\delta (\omega + \mu + |U|) + \left( 1 - \frac{n}{2} \right)
\delta (\omega + \mu) , 
\label{nine}
\end{equation}
from which the following properties are inferred.
(i) The spectral function has the form of two peaks with 
weights $\frac{n}{2}$ and $1-\frac{n}{2}$ (which are very
different if $n \ll 1$). (ii) The two peaks are separated by the binding
energy of the pairs (which is $|U|$ in the atomic limit).
(iii) At zero temperature, $\mu = -\frac{|U|}{2}$, and in
Eq. (\ref{eight}) the Fermi-function eliminates the second
peak of $A({\bf k}, \omega)$. The resulting intensity
$I({\bf k}, \omega)$ is a single peak located $\frac{|U|}{2}$
below the chemical potential. The system would therefore
display a ``pseudogap" (a true gap in this case) of size $\frac{|U|}{2}$.
(iv) At higher temperatures, the Fermi-function is smoothed out 
giving rise to the second peak at a higher energy and weakening the
first one, so the former might become stronger than the latter.

In the general case of finite $t/|U|$, non-zero kinetic energy 
leads to
a number of new effects. It reduces the binding energy, i.e.
the interpeak distance, and assigns finite widths to the peaks
of $A({\bf k},\omega)$. Next, it restores
the ${\bf k}$-dependence of $A({\bf k},\omega)$ and $I({\bf k},\omega)$.
Finally, due to the finite radii of pairs and their ovelapping,
the binding energy becomes density-dependent \cite{Kyung}.
However, our numerical results show that properties (i)-(iv) 
listed above remain valid even at finite $t/|U|$. 
Moreover, we believe they are generic to any
fermion model with attraction in the low-density limit.

\begin{figure}[t]
\begin{center}
\leavevmode
\hbox{
\epsfxsize=8.6cm
\epsffile{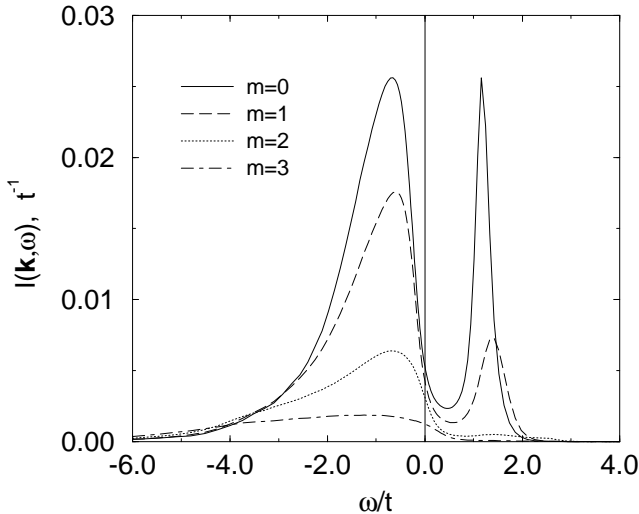}
}
\end{center}
\vspace{-0.5cm}
\caption{ 
The intensity of ARPES $I({\bf k},\omega) = A({\bf k},\omega) f_F(\omega)$
for density $n=0.017$, temperature $T=0.3\,t$, and several
momenta ${\bf k}=m(\frac{\pi}{8},\frac{\pi}{8})$, $m=0,1,2,$
and 3 (from the top curve down).
}
\label{fig1}
\end{figure}

In Fig. \ref{fig1} we show the solution of
Eqs. (\ref{two})-(\ref{eight}) 
for the lowest density $n=0.017$ and $T=0.3 \, t$.
$I({\bf k},\omega)$ displays a complicated ${\bf k}$-
and $\omega$-dependence which can be understood as follows.
A particle with momentum ${\bf k}$ can be found in two
distinctly different states: either in a state of the 
single-particle band with an extended wave function or as
a component of a bound state with a localized wave function. The
two possibilities give rise to $A({\bf k},\omega)$, which consists
of two peaks separated by the pair binding energy 
$\triangle E = 2.1\, t$ (for $|U|=8\,t$).   
Multiplication by the Fermi-function
cuts off the high-energy (single-particle band)
peak, which depends on the energy of the latter and the 
temperature. For ${\bf k} = (0,0)$ in Fig. \ref{fig1} the 
high-energy peak is reduced in height significantly --- to that of 
the low-energy peak, but not to zero. Note that after the
cut-off, the peak is slightly shifted from its original position.
The position of the high-energy peak disperses with 
{\bf k} as does the bare specrum $\varepsilon_{\bf k}$,
and, as ${\bf k}$ increases, the peak gets cut off by the 
Fermi-function very rapidly (compare the cases for the
different momenta ${\bf k}$ 
in Fig. \ref{fig1} for $\omega > 0$). Let us now turn to
the low-energy peak. The probability of finding a particle
with momentum ${\bf k}$ in a bound state is the square
of the bound state's wave function. For zero total momentum
${\bf P}$, one has
\begin{equation}
\psi_{{\bf P}=(0,0)}({\bf k}) = \frac{C}{E-2\, \varepsilon({\bf k})},
\label{ten}
\end{equation}
where $C$ is the normalization constant and $E$ the energy of the
bound state measured from the bare atomic level. 
The relative height of low-energy peaks in Fig. \ref{fig1} is
in good agreement with $|\psi({\bf k})|^2$
for $E= 2\, \varepsilon(0,0) - \triangle E = -10.1 \, t$.
This corroborates the bound-state origin of the low energy peaks
in $A({\bf k},\omega)$ and $I({\bf k},\omega)$.
Thus, on the basis of Fig. \ref{fig1}, we conclude that
ARPES of the attractive Hubbard model exhibit
a clear pseudogap behaviour at low temperatures and densities.
The momentum and frequency dependences of the spectra
have simple physical explanations.

\begin{figure}[t]
\begin{center}
\leavevmode
\hbox{
\epsfxsize=8.6cm 
\epsffile{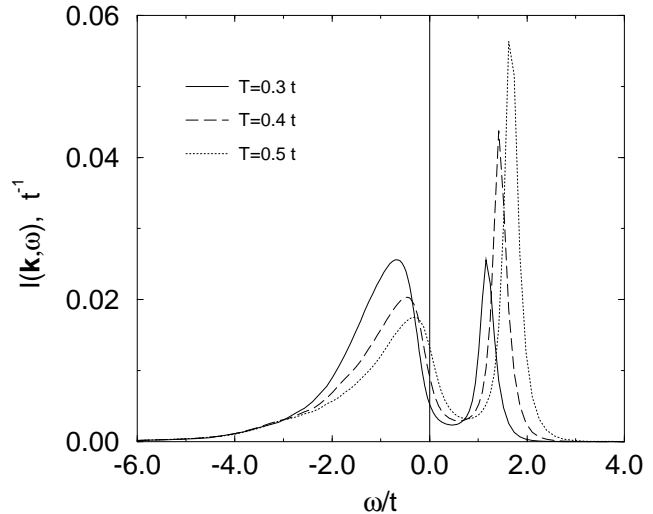}
}
\end{center}
\vspace{-0.5cm}
\caption{
The temperature dependence of $I({\bf k},\omega)$ for
density $n=0.017$ and momentum ${\bf k}=(0,0)$.
}
\label{fig2}
\end{figure}

The temperature dependence of $I({\bf k},\omega)$ for $n=0.017$ and
${\bf k}=(0,0)$ is shown in Fig. \ref{fig2}. There are two main effects as
temperature increases. First, weight is transferred from
the low-energy peak to the high-energy one. This is due to the
progressive thermal excitation of particles to the single-particle
band and consequent weaker influence of the bound states on the
single-particle spectral function. 
Secondly, the whole structure moves to
higher energies relative to the chemical potential. These
two effects lead to the rapid suppression of the pseudogap
as temperature increases. Note that the distance between
the two peaks is $T$-independent and remains approximately
the pair binding energy (slightly reduced by the cut off), 
in accordance with the atomic limit.
\begin{figure}[t]
\begin{center}
\leavevmode
\hbox{
\epsfxsize=8.6cm 
\epsffile{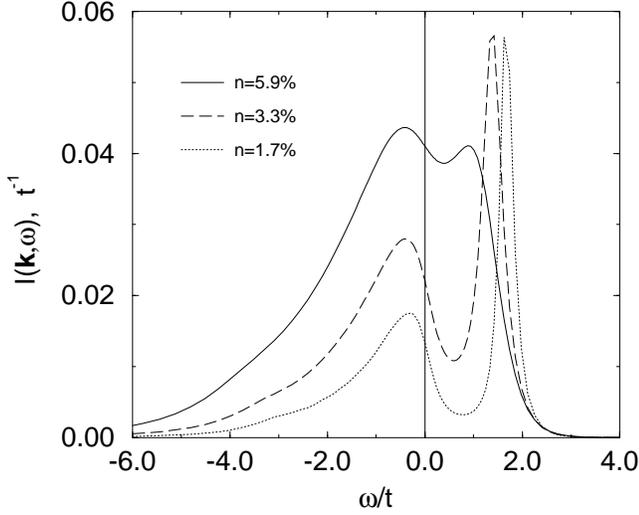}
}
\end{center}
\vspace{-0.5cm}
\caption{
The density dependence of $I({\bf k}, \omega)$ for temperature
$T=0.5\,t$ and momentum ${\bf k}=(0,0)$.
}
\label{fig3}
\end{figure}

Fig. \ref{fig3} presents the density dependence of $I({\bf k}, \omega)$
for $T=0.5\,t$ and ${\bf k}=(0,0)$. Clearly, the pseudogap
disappears as $n$ increases. We have already argued
elsewhere \cite{Kyung} that this is a result of the rise
of the two-particle level due to the packing effect when
pairs begin to overlap. (Intuitive arguments of this kind
were given earlier in \cite{Randeria_one}.) 
Since the binding energy decreases
with $n$, the temperature becomes progressively more effective 
in unbounding pairs, washing away the pseudogap.

It is quite remarkable that such a simple system as the
attractive Hubbard model and such complex systems as 
high-$T_c$ superconductors have very similar dynamical
properties. They both display pseudogaps at low temperatures
and carrier densities, which disappear as $T$ and $n$ increase.
This suggests the conclusion that the carriers in HTSC do
experience some sort of short-range attraction. HTSC 
therefore exhibit properties which are generic to fermionic
systems with attraction and which are captured in our
model calculations.

\begin{figure}[t]
\begin{center}
\leavevmode
\hbox{
\epsfxsize=8.6cm
\epsffile{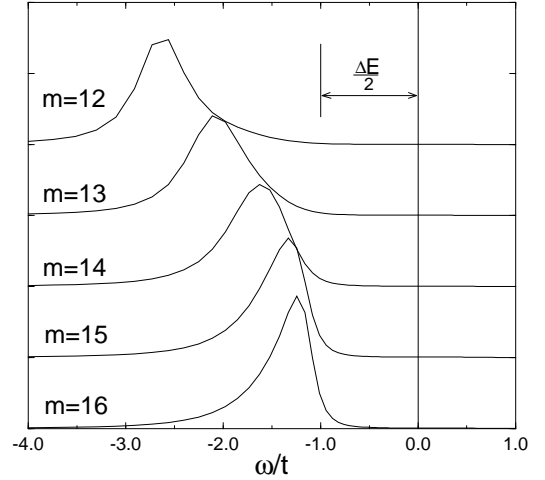}
}
\end{center}
\vspace{-0.5cm}
\caption{
$I({\bf k}, \omega)$ for the nearly complete filling $n=1.983$,
temperature $T=0.3\,t$, and several momenta 
${\bf k}=m\,(\frac{\pi}{16},\frac{\pi}{16})$. 
}
\label{fig4}
\end{figure}

One could now proceed in elaborating the model while
trying to keep the properties obtained intact. Further
insight into the problem can be gained by considering
the opposite limit of {\em nearly complete} filling $2-n \ll 1$.
In this case, Eq.~(\ref{one}) may be viewed as a phenomenological
model for {\em electrons} rather than holes. 
(To some extent, the nearly fully filled band imitates the
nearly filled lower Hubbard band when the Coulomb
repulsion is taken into account. Unfortunately, this analogy 
is not complete, due to different temperature behaviour 
of $\mu$; see below.) The quantity
$I({\bf k}, \omega)$ of Eq. (\ref{eight}) has now the
meaning of the number of electrons emitted from the
system per time unit, which brings the whole model
closer to reality. 
There is no need to recalculate the spectra, since on a bipartite 
lattice the dilute and nearly filled limits are related
by the particle-hole transformation, which leads to the relation 
\begin{equation}
A({\bf k}, \omega ; n) f_F(\omega) =
A({\bf k} + {\bf Q}, -\omega ; 2-n) [1 - f_F(-\omega)],
\label{eleven}
\end{equation}
where ${\bf Q}=(\pi,\pi)$ for the square lattice. Due to
the inversion of the occupation numbers, it is now the
pairing-induced peak that gets cut off by  
Fermi-function. The inversion of the frequency places the
remaining peak below the chemical potential. The resulting
ARPES are shown in Fig. \ref{fig4}. They are
complementary to the spectra of Fig. \ref{fig1}. Clearly, pseudogap
is present, because the spectrum with the largest momentum 
${\bf k}=(\pi,\pi)$ is still peaked far below the chemical 
potential.  

The overall picture looks very much like spectra of
a weakly-interacting system {\em but shifted} from $\mu$ by
half of the pair binding energy. We emphasize that
pairs themselves are {\em not} seen explicitly in the
spectra, since the pair-induced peak of $A({\bf k},\omega)$
has been cut off by the Fermi-function. Nevertheless, the
pairs are present implicitly, manifesting themselves
in the shift of the chemical potential. This observation
is important for understanding the ARPES of HTSC.

We do not present temperature and density dependences of
$I({\bf k},\omega)$ for the nearly filled case, for they
are complementary to Figs. \ref{fig2} and \ref{fig3},
respectively. The pseudogap now vanishes with {\em decreasing}
electron density (increasing hole doping), in accordance
with Fig. \ref{fig3}. The temperature dependence is, however,
different from the dilute limit. At nearly full filling,
the chemical potential {\em goes up}, with temperature
and the distance between $\mu$ and the single-particle
band increases. Therefore, the pseudogap is expected to rise
with $T$ in this case. To obtain the correct temperature
behaviour one would need to consider the density regime
close to half-filling, which is outside the range of validity of
the $T$-matrix approximation.

In conclusion, we have shown that photoemission spectra
of the attractive Hubbard model in the low-density limit
display a clear pseudogap behavior, qualitatively similar
to that of high-$T_c$ superconductors. Our findings support
the suggestion that the pseudogap feature observed in HTSC 
results from real-space pairing and the formation of bound 
pairs in the normal state. 

We thank E.\,G.\,Klepfish for valuable discussions on the
subject. PEK acknowledges the support of EPSRC grant
GR/L40113.

\end{document}